\documentclass[aps,pra,reprint,superscriptaddress,showkeys]{revtex4-1}
\usepackage{graphicx}
\usepackage{bm}
\usepackage{amsmath,amssymb}
\usepackage{epsfig}
\usepackage{ booktabs} 
\usepackage{epstopdf}
\usepackage{dsfont}
\usepackage{color}
\usepackage{changes}
\usepackage{float}
\usepackage{dcolumn}

\newcommand{\red}[1]{\textcolor{black}{#1}}
\newcommand{\bnew}[1]{\textcolor{black}{#1}}

\newcommand{\xm}[0]{\langle x_{min} \rangle}
\newcommand{\xx}{{ x_{max}} }
\newcommand{\tagalign}[1]{\stepcounter{equation}\tag{\theequation}{#1}}
\newcommand{\tens}[1]{\underline{\underline{#1}}}
\newcommand{\vc}[1]{\underline{#1}}
\newcommand{\tildet}{\tilde{\theta}}
\newcommand{\dup}{\mathrm{d}}
\newcommand{\tightint}[2]{\hspace{-1em}\int\displaylimits_{#1}^{#2}\hspace{-1em}}

\begin{document}

\preprint{APS/123-QED}

\title{Signatures of the spatial extent of plastic events in the yielding transition in amorphous solids}

\author{Daniel Korchinski}
 \affiliation{Department of Physics and Astronomy and Stewart Blusson Quantum Matter Institute, University of British Columbia, Vancouver BC V6T 1Z1, Canada}
 
\author{C\'eline Ruscher}%
 \email{celine.ruscher@ics-cnrs.unistra.fr}
\affiliation{Institut Charles Sadron, 23 rue du Loess, F-67034 Strasbourg, France}

\author{J\"org Rottler}
 \affiliation{Department of Physics and Astronomy and Stewart Blusson Quantum Matter Institute, University of British Columbia, Vancouver BC V6T 1Z1, Canada}

\date{\today}

\begin{abstract}
Amorphous solids are yield stress materials that flow when a sufficient load is applied. Their flow consists of periods of elastic loading interrupted by rapid stress drops, or avalanches, coming from microscopic rearrangements known as shear transformations (STs). Here we show that the spatial extent of avalanches in a steadily sheared amorphous solid has a profound effect on the distribution of local residual stresses $x$. We find that in this distribution, the most unstable sites are located in a system size dependent plateau. While the entrance into the plateau is set by the lower cutoff of the mechanical noise produced by individual STs, the departure from the usually assumed power-law (pseudogap) form $P(x) \sim x^{\theta}$ comes from far field effects related to spatially extended rearrangements. Interestingly, we observe that the average value of weakest sites $\xm$ is located in an intermediate power law regime between the pseudogap and the plateau regimes, whose exponent decreases with system size. Our findings imply a new scaling relation linking the exponents characterizing the avalanche size and residual stress distributions. 

\end{abstract}

\maketitle

\section{Introduction}

Upon deformation, amorphous materials behave as solids when the applied shear stress is lower than the yield stress and start to flow when this threshold is exceeded. 
In the limit of very slow shear rate and at low temperature, the stress response becomes very jerky as seen for instance in bulk metallic glasses \cite{hufnagel2016deformation,zhang2017expe},  foams \cite{Cantat2006}, granular matter \cite{Dahmen2011} or porous silica \cite{Baro2013}.  The sudden stress drops, or avalanches, originate in localized microscopic plastic rearrangements involving a small number of particles in shear transformation zones (STs) \cite{Argon1979,spaepen1977}, which have been observed both in atomistic simulations \cite{FalkLanger1998} and in colloidal glasses with confocal microscopy \cite{Schall2007}. Avalanches of size $S$ are expected to be scale free in the thermodynamic limit following the distribution $P(S) \sim S^{-\tau}$ where $\tau $ is the avalanche exponent. However, for finite systems of linear dimension $L$, they present an upper cutoff $S_c \sim L^{d_f}$ where $d_f$ is the fractal dimension that characterises the geometry of the failure event.
\begin{figure}
     \centering
     \includegraphics{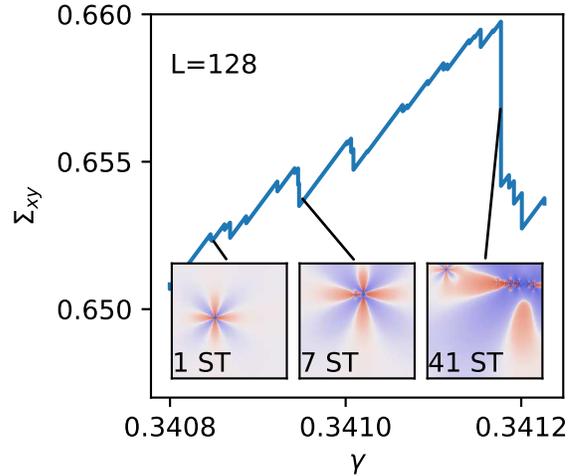}
     \caption{Segment of a stress-strain obtained from the elastoplastic model with $L=128$. Stress drops correspond to plastic rearrangements comprised of shear transformations. Insets: Plastic activity in the model for different sizes of stress-drop. Larger avalanches tend to form line-like events.   \label{fig:epm_behaviour} }
 \end{figure}

\red{Our present understanding of the yielding phenomenon in these materials is built upon the notion of anisotropic elastic interactions produced by individual STs in the surrounding medium \cite{Eshelby1957}. Since these interactions are long-ranged with an alternating sign, they stabilize or destabilize distant sites, and therefore act as a mechanical noise. As the load $\Sigma$ approaches the yield (or critical) stress $\Sigma_Y$, however, plastic events become collective and strongly correlated. The presence of such correlations is signaled, for instance by anomalous scaling of the fluctuations of the stress, $\delta \Sigma \sim L^{-\phi}$, where $L$ is the linear size of a system of dimension $d$ \cite{Karmakar2010,SalernoRobbins,LinWyart2014}.  Multiple studies have consistently found $\phi < d/2$ at criticality, i.e.~stresses do not add up independently but are correlated.
The mechanical noise bath experienced by a distant site therefore not only contains contributions from single, point-like STs, but also from extended, line-like plastic events. While the stress change due to a distant point-like event is of order $L^{-d}$, the typical one induced by an extended avalanche is of order $S_c/L^d\sim L^{-(d-df)}$, where $d_f\le d$ in amorphous materials.}

In this work, we first show that the stress fluctuations produced by these spatially extended plastic events have distinctly different statistical properties than those coming from individual STs. We then proceed by showing that the presence of the collective avalanches has a profound and so far overlooked effect on the stability of very slowly deformed amorphous solids. In the limit when thermal effects play no role, often referred to as the athermal quasistatic regime (AQS), the statistical properties of the macroscopic failure events (avalanches) are closely linked to the distribution of residual stresses, i.e.~how far a given local region finds itself from instability. This distribution is very sensitive to the underlying mechanical noise. Indeed we will show that there exists a transition that arises due to the mechanical noise associated to spatially extended events and that this transition occurs above the average value of weakest site.

As a consequence, a central result of our work is that the spatial shape of the plastic events, reflected in their fractal dimension $d_f$, enters the system size scaling of the typical value of the weakest site, which in turn controls the mean stress drop caused by avalanches under steady state flow conditions. We then propose a new scaling law linking the exponents characterizing the avalanche size and residual stress distributions. While a numerical demonstration of these concepts can only be performed for finite system sizes, we argue that the influence of the spatial extent of the avalanches remains in the thermodynamic limit.

\section{Elastoplastic Model}
 \red{We use a coarse-grained, elastoplastic mesoscopic model (EPM) \cite{nicolas2018deformation}, which contains $N=L^2$ sites on a two-dimensional square lattice, that is ideally suited to study the statistics of avalanches and residual stresses. In our finite element implementation \cite{budrikis2017universal}, the Eshelby stress propagator $\mathcal{G}(r) \sim \cos(4\theta)/r^d$ \cite{Eshelby1957,Picard2004} for STs emerges naturally.  Sites yield when the local stress exceeds their local yield threshold $|\sigma_{xy}| > \sigma_Y$. Upon yielding, the sites reduce their local stress to zero by accumulating plastic strain.  The yield threshold $\sigma_Y$ for that site is then redrawn from a Weibull distribution with shape parameter $k=2$, the same distribution that is used to initialize the yield thresholds. }
 
\red{We implement a strain-controlled deformation protocol using extremal dynamics  \cite{Talamali2011}, in which avalanches are initiated by uniformly loading the system in a simple shear configuration until the weakest site fails, and fixing the strain until the avalanche ends (cf. Fig.~\ref{fig:epm_behaviour}). Single STs have the characteristic quadrupolar form, while large avalanches show a line-like structure (cf. inset of Fig~\ref{fig:epm_behaviour}). Loading is done by specifying the displacement at the boundary vertices of the system. We thus use surfaces instead of periodic boundary conditions \cite{sandfeld2015avalanches}. Previous studies indicate that the universal critical behavior is insensitive to the detailed form of the loading conditions \cite{Budrikis2017}. From a stable configuration, the system is loaded to until one site exactly meets its yield stress: $|\sigma_{xy}| = \sigma_Y$. The yielding site initiates an avalanche. After a site yields, all sites are checked for stability. The most unstable site then yields next, until the system returns to stability. Throughout the avalanche, the displacements at the boundaries of the system remain fixed. All data reported in this paper was taken after a steady-state flow state was reached (initial transients were discarded).}

 \begin{figure*}[t]
\centering
\includegraphics{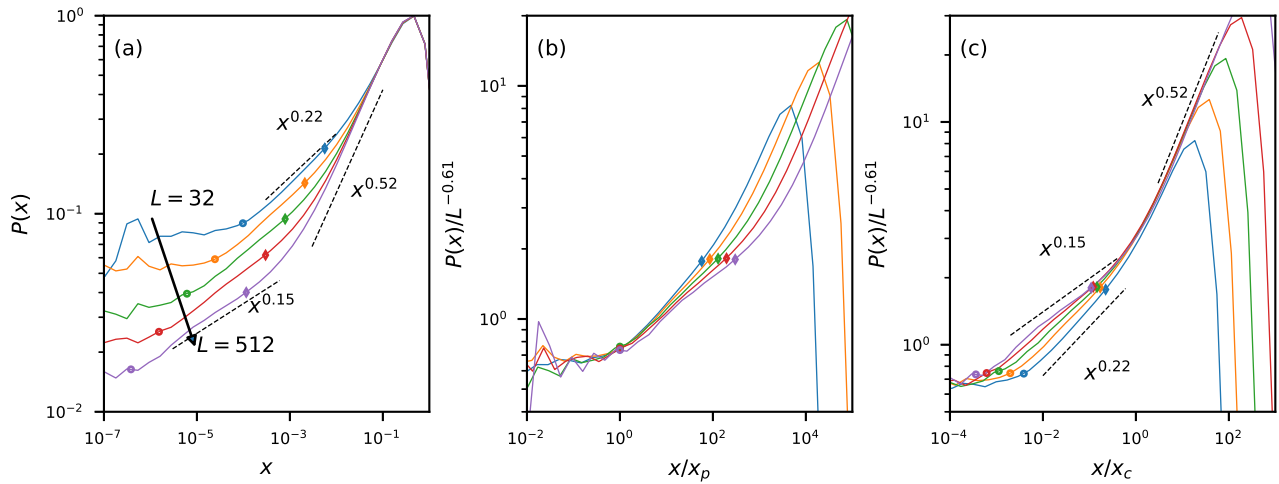}
\caption{(a) Probability distribution function of the residual stresses $P(x)$ for different system sizes $L$. In panels (b) and (c) $x$ has been rescaled by the $x_p\sim L^{-2.0}$ and by $x_c \sim L^{-1.15}$ (see text and SI Fig.~\ref{fig:fss}), respectively. Filled circles indicate the location of the lower cutoff of the mechanical noise $\delta x_c\sim L^{-2.0}$ and filled diamonds indicate the mean values of the weakest site $\xm$.
\label{fig1}}
\end{figure*}

\section{Distribution of residual stresses}
The probability distribution function $P(x)$ of residual stresses $x=\sigma_Y-\sigma_{xy}$ is shown in Fig \ref{fig1}(a) for different system sizes and exhibits three distinct regimes. For larger values of $x$ we observe that $P(x) \sim x^{\theta}$ with a pseudogap exponent $\theta \approx 0.5$ in good agreement the literature \cite{LinWyart2014,LiuBarrat2016,Budrikis2017}. \red{Several recent studies have investigated the behavior of $P(x)$ for smaller values of $x$ in more detail, and reported that the power law regime gives way to a finite plateau value as $x\rightarrow 0$ \cite{Tyukodi2019,Ferrero2019,Ruscher2020}. Our results here confirm a departure from the pseudogap regime, but reveal that the situation is more complex:  Below a crossover value $x_c$, $P(x)$ in fact develops an intermediate power law regime $P(x) \sim x^{\tilde{\theta}}$ with $\tilde{\theta}<\theta$, before  finally saturating in a system size dependent plateau value $P_0$ for $x \rightarrow 0$. }

\subsection{Origin of the terminal plateau}
To gain more insight into what happens for $x<x_c$, we first investigate the origin of the plateau region. As already observed in ref.~\cite{Tyukodi2019}, the plateau depends on the system size as $P_0 \sim L^{-p}$. We find that $p \approx 0.61$  (Appendix Fig.~\ref{fig:fss}) as also reported in \cite{Tyukodi2019, Ferrero2019}. In ref.~\cite{Ruscher2020} it was suggested that the emergence of the plateau is related to the discreteness of the underlying mechanical noise arising from the stress redistribution during avalanches. For a given site, one defines the mechanical noise $\delta x_i = x_i(n+1)-x_i(n)$ where $n$ is the number of plastic events in chronological order. Due to the long-range nature of the elastic interaction described by the stress propagator $\mathcal{G}(r) \sim r^{-s}$, this noise is broadly distributed \cite{Lemaitre2007} and can be expected to follow a L\'evy distribution $P (\delta x) \sim |\delta x|^{-\mu -1}$ \cite{LinWyartPRX}.  The aforementioned exponent $s=d/\mu$ and when $\mu=1$, $\mathcal{G}(r)$ corresponds to the Eshelby propagator for STs. Assuming the different rearrangements are independent and correspond to single STs, one expects a lower cutoff $\delta x_c \sim L^{-d/\mu}$ for the noise distribution. The definition of the lower cutoff is not unique \cite{Ferrero2020, ParleySollich2020} and depends on the nature of the plastic objects considered (single STs vs. large avalanches). This aspect will be discussed in more detail below. The noise distribution also presents an upper cutoff $\delta x_{uc} \sim 1$ coming from the yield events occurring at neighboring sites. 

\begin{figure}[t]
\centering
    \includegraphics[]{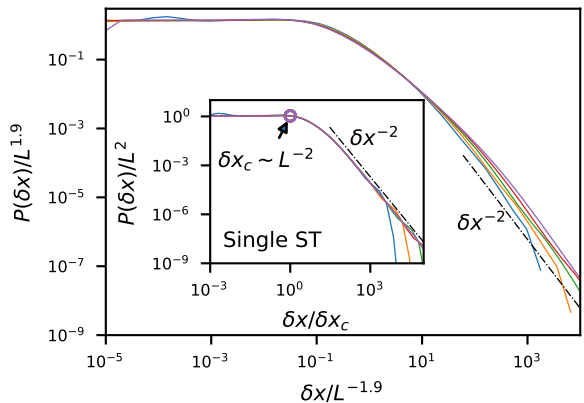}
\caption{Finite-size scaling of the mechanical kick distributions experienced by sites over the course of plastic rearrangements. Inset: Kick distributions from plastic events with a single ST, with the onset of the plateau defining $\delta x_c$.  
	\label{fig2}}
\end{figure}

In Fig.~\ref{fig2} we show the distributions of stress changes $\delta x$, also called ``kicks'',  for plastic activity measured in our EPM for different system sizes. These distributions collapse when the stress change is rescaled by \bnew{$L^{-1.9}$} and the distribution scales as $P(\delta x) \sim \delta x^{-2}$, \bnew{which is very nearly the same as for the single ST Eshelby interactions $(\mu=1)$ we show in the inset.} We conclude that the overall noise mechanical noise is largely dominated by point-like events. 
\bnew{These single STs set the smallest scale in the system, with a lower cutoff kick-size $\delta x_c$.} To highlight the role of the lower cutoff in the emergence of the plateau in $P(x)$, we use $x_p = \delta x_c$ and consider a rescaled distribution $P(x)L^{-p}$ vs. $x/\delta x_c$ in Fig \ref{fig1}(b). A good finite system size collapse is obtained for $x$ below and in the vicinity of  $\delta x_c$ showing that this region is dominated by the influence of the lower cutoff of the mechanical noise $\delta x_c$.

\subsection{Origin of the intermediate regime}
However, this rescaling fails to collapse points at or above $\xm$, the average value of the weakest site and the second smallest characteristic scale indicated by diamonds in Fig \ref{fig1}. We notice that values of $\xm$ are systematically located below the crossover $x \lesssim x_c$ that demarks the departure from the pseudogap regime $P(x)\sim x^\theta$ and is defined as the intersection between the two power law regimes. This crossover scales as $x_c \sim L^{-c}$ with $c \approx 1.15$ (Appendix Fig.~\ref{fig:fss}). In  Fig.~\ref{fig1}(c), we show $P(x)L^{0.61}$ vs $x/x_c$ and find a good collapse of the upper power law regime in the region $x > x_c$ and of the plateau region. However, in the intermediate region where $\delta x_c < x< x_c$ the collapse fails. The phenomenology below and above $x_c$ is thus different as suggested by the different scaling with system size.

\begin{figure}[t]
\centering
    \includegraphics[]{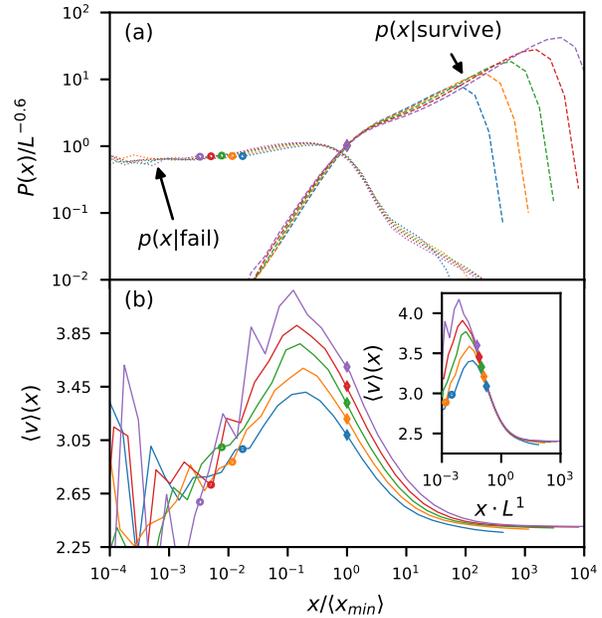}
	\caption{(a) Representation of the distribution $P(x)$ rescaled by $\xm$ conditioned on survival and absorption in the next avalanche. (b) Finite-size evolution of the drift $v$ experienced by sites with respect to the value of residual stresses. Inset: $\langle v \rangle(x)$, rescaled by $x\cdot L$, showing that the onset of drift enhancement occurs with a different scaling than the maximum drift enhancement.}
	\label{fig3}
\end{figure}

To shed more light on the origin of this intermediate power law regime, we focus our attention on $\xm$. It is intrinsically related to the macroscopic flow, in particular the average value of stress drops $\langle |\Delta \sigma| \rangle \sim \xm \sim L^{-\alpha}$, where $\alpha=1.4$ (Appendix Fig.~\ref{fig:fss}).  One can envision $P(x)$ as the survival probability of a random walker performing a L\'evy flight near an absorbing boundary \cite{LinWyartPRX}. It is thus interesting to investigate $P(x)$ subject to the condition of absorption or survival after the occurrence of an avalanche. Results are shown in Fig \ref{fig3}(a), where we see first that, as mentioned above, no site survives below $\delta x_c$ and most of the sites in the plateau region are likely to be absorbed after an avalanche. We then observe that $P(x)$ conditioned on the surviving sites departs from the pseudogap regime when $x \approx  \xm $, where both failing and surviving distributions are equal. Therefore, $\xm$ marks the onset of a transition; for $x > \xm$ sites are more likely to survive while below $\xm$,  absorption dominates.
Moreover, the extremal dynamics protocol induces a global shift $-x_{min}$ of residual stresses to initiate each avalanche. This invites the introduction of a drift velocity:  $ v =N x_{min} / \ell$ over the course of an avalanche with $\ell$ plastic events. \bnew{ An interesting question is then: what drift do sites experience before landing at a particular value of $x$? We check this in Fig.~\ref{fig3}b, and find a strong size-dependent enhancement in the drift velocity.}
For a L\'evy flight of index $\mu=1$ with a drift $v$, the persistence exponent, i.e. the pseudogap exponent in the context of sheared amorphous solids, can be expressed as a function $v$ and the amplitude of the mechanical noise $A$ as $\theta = \arctan(A\pi/v)/\pi$ \cite{Doussal2009, LinWyartPRX}. Assuming $A$ to be constant, one expects a decrease of $\theta$ with increasing $v$ that we compute in the regions $x > \xm$ and $x < \xm$ in Fig \ref{fig3}(b). The inset reveals that, although the maximum drift scales as $\xm$, the drift enhancement begins at $x \sim L^{-1} \approx x_c$. This observation is coherent with the increase in probability of failing shown above. 

\begin{figure}[t]
\centering
    \includegraphics[]{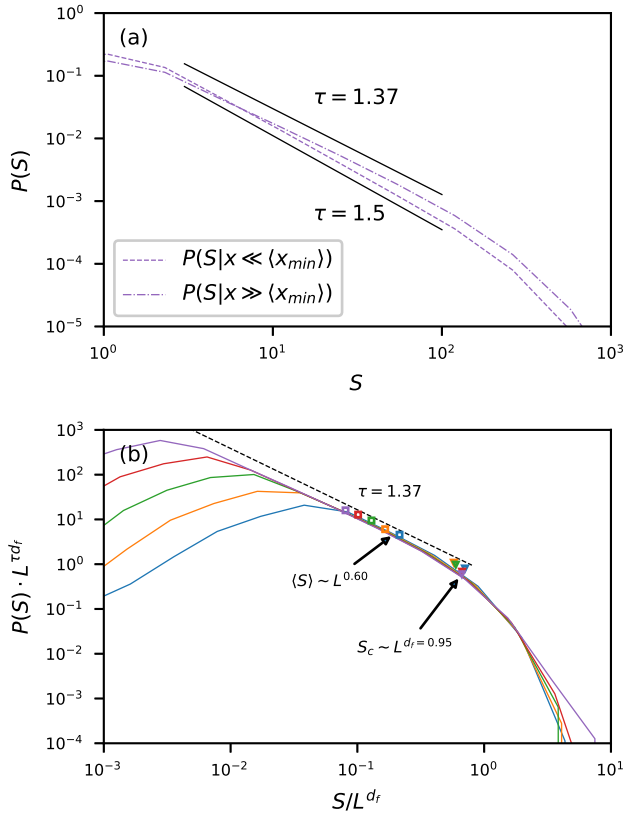}
	\caption{(a) Avalanche size distribution for $L=512$ conditioned on $x < \xm$ and $x > \xm$. (b) Unconditioned avalanche distributions $P(S,L)$. Colors indicate system sizes as in Fig.~\ref{fig1}.
	} 
	\label{fig4}
\end{figure}

The drift also reveals a change in the nature of avalanches above and below $\xm$. Indeed $\tilde{v} > v$ implies that $\langle\tilde{\ell}\rangle < \langle \ell \rangle$. The typical avalanches are larger above $\xm$, which thus marks a transition between collective and individual rearrangements. To illustrate this difference\bnew{, we study which avalanches lead to sites much less or much more stable than $ \xm$. This is accomplished by, for each avalanche of size $S$ and for each site $l$ with stability $x_l$, constructing the pairs $(x_l, S)$. For those pairs with $x_l \ll \xm$ (or $\gg \xm$), we compute the distribution of $S$ as $P(S | x \ll \xm)$ (or $P(S|x \gg \xm)$)} which are shown in Fig \ref{fig4}(a). We emphasize that even if we focus on $ x \ll \xm$, we observe large values of $S$ as they can lead to values of $x < \xm$. We notice that the avalanche exponent $\tau$ differs substantially. Indeed, for sites well above $\xm$, $\tau \approx 1.37$ corresponds to the avalanche exponent measured for the whole distribution while for $x \ll \xm$, $\tau \approx 1.5$ as expected in a mean-field picture of plasticity emerging from rearrangements of independent STs. Interestingly, the same value of $\tau$ has been measured for small avalanches in atomistic simulations \cite{Oyama2020}. Our interpretation is also consistent with the findings of Karimi et al. \cite{karimi2017}, who reported much narrower distributions of inertial (extended) avalanches than for overdamped (localized) ones. 

The decrease of $\tilde{\theta}$ with increasing system size $L$ (see also Appendix Fig.~\ref{fig:fss}) comes from the joint effect of the decrease in the density of sites below $\xm$ with system size as the cumulative distribution function $\int_0^{\xm} P(x)dx \sim L^{-d}$ \cite{Karmakar2010}, and of an increase of the drift $\tilde{v}$ with $L$ as observed  in Fig \ref{fig3}(b). Assuming that the average of failing sites per avalanche is $\langle \tilde{\ell} \rangle \approx 1$, the drift $\tilde{v} \sim \langle S \rangle \sim L^{d-\alpha}$ and therefore we immediately see that the larger $L$ the larger the drift in the region where $x < \xm$. More and more sites are brought on the verge of instability and in the thermodynamic limit, when $L\rightarrow \infty$, one expects the drift to become infinite and the density of sites below $\xm$ to be zero. This implies $\tilde{\theta}(L) \rightarrow 0$ and thus $P(x)$ should plateau for $x  \le \xm$.

From the discussion so far, we hypothesise that the probability of residual stresses can be written as 
\begin{equation}
P(x;L)= \left\lbrace
\begin{array}{ll}
	& P_0(L) \hspace{1.0cm} \forall \,  \,  x \le  x_p 	 \\
    & c_1(L) x^{\tilde{\theta}(L)} \hspace{0.3cm} \forall  \,   \,   x_p	 \le x \le x_c \\
	& c_2 x^{\theta} \hspace{1.2cm} \forall  \,   \,  x \ge x_c
	\label{pdex_eq} 
\end{array}
\right. 
\end{equation}
Continuity at $x=x_p$ implies $c_1(L) \sim L^{2\tilde{\theta}-p}$. In the thermodynamic limit, $P(x)$ below $x_c$ reduces to $P(x) = \tilde{P_0}$ where $\tilde{P_0} \sim L^{-p}$. Moreover, the continuity at $x=x_c$ and the relation $x_c \sim L^{-c}$ allow us to establish an expression for the plateau exponent,
\begin{align}
	 p=c \theta + \tilde{\theta}(L)(2 -c)  \stackrel{L \rightarrow \infty}{=} c \theta 
	 \label{eq_for_p}
\end{align}
In the thermodynamic limit, when $\tilde{\theta}$ vanishes, with $c \approx 1.15$ and $\theta\approx 0.52$, this relation predicts $p = 0.60$ in good agreement with the value we find meaning that the intermediate power law has not strong influence on $p$ for the range of sizes investigated.

\begin{figure}[t]
\centering 
\includegraphics[]{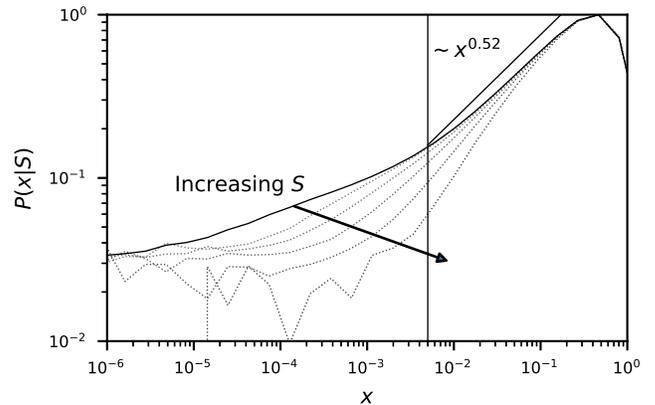}
\caption{ Solid line is the $P(x)$ distribution for L=256. Dashed lines are $P(x)$ immediately after large avalanches of increasing size. 
}
\label{fig:px_s}
\end{figure}

\section{Influence of the spatial extent of avalanches}
\subsection{Origin of the crossover $P(x)$}
For now we have linked the emergence of the plateau in $P(x)$ to the lower cutoff of mechanical noise and showed that $\xm$ sets the scale \red{at which the distributions $P(x)$ conditioned on failing vs. surviving sites become equal. However, it is evident from Fig.~\ref{fig3}(b) that the drift enhancement begins already for $x>\xm$, which is also signaled by a sharp increase in $P(x|{\rm fail})$ in the same region. These observations point to the presence of an additional scale.}
The question of the origin of the crossover in $P(x)$ remains as well. What is the physical meaning of $x_c$? To tackle this question, we investigate the role of the spatial extent of avalanches on the distributions of residual stresses and mechanical noise. In practice, we adopt a top to bottom approach by considering only specific sizes of avalanches and by conditioning the probability distributions on these specific values.

As a first step, we compute the unconditional distribution of avalanches $P(S)$ and  find in Fig \ref{fig4}(b) an avalanche exponent $\tau \approx 1.37$ and a fractal dimension $d_f \approx 0.95$ consistent with results from atomistic simulations \cite{LiuBarrat2016, SalernoRobbins} but slightly lower than results from other EPM implementations \cite{LinWyart2014,Ferrero2019}. \bnew{In analogy to Fig.~\ref{fig4}(a), we consider the distribution of residual stresses in the system, conditioned on the preceding avalanche and find in Fig.~\mbox{\ref{fig:px_s}} a plateau that depends on the size of the avalanche. Since $P(x) = \int \dup s P(x|s)P(s)$, the plateau with the earliest (i.e. largest $x$) onset in $P(x|s)$ must therefore result in the deflection from the power-law.} 

\begin{figure}[t]
\centering
    \includegraphics[]{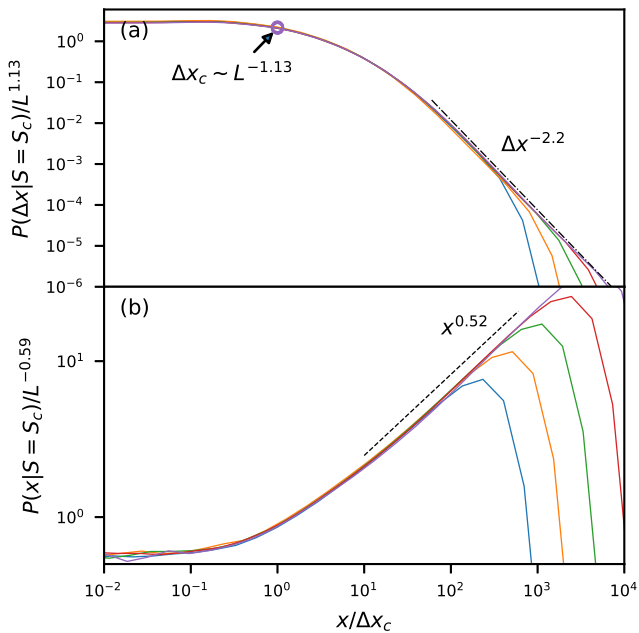}
	\caption{(a) Kick distribution from avalanches $S\approx S_c$, with finite-size scaling effected. (b) Residual stress distribution after avalanches with size $S\approx S_c$, collapsed with the same finite-size scaling. }
	\label{fig5}
\end{figure}

\bnew{The typical stress release of the largest avalanches is set by $S_c$, corresponding to system-spanning events. Avalanches of this size would correspond to the scale of the earliest deviation from $P(x)\sim x^\theta$ at $x_c$. Fig.~\mbox{\ref{fig5}} tests this hypothesis by considering the distribution of stress kicks $\Delta x$ on sites from plastic events of size $S\approx S_c$. 
As can be seen in Fig.~\mbox{\ref{fig5}}a), this changes the power law regime $P(\delta x)\sim \delta x^{-2.0}$ found for all kicks to $P(\Delta x) \sim \Delta x^{-2.2}$ and implies an apparent value of $\mu=1.2$. The typical elastic interaction can thus be seen as more long-ranged \cite{ParleySollich2020} with an effective kernel decaying as $\mathcal{G}(r) \sim r^{-1.8}$. Moreover, the lower cutoff of the kick distribution now scales as $\Delta x_c \sim L^{-1.13}$ (Fig.~\mbox{\ref{fig5}}a), which has a finite-size scaling in near perfect agreement with $x_c \sim L^{-1.15}$ found from direct analysis of $P(x)$ (Fig.~\mbox{\ref{fig1}}c and Appendix Fig.~\mbox{\ref{fig:fss}}). This behavior reflects collective instabilities, where one unstable site triggers further rearrangements and so on. A distant site not only feels one rearrangement but an apparent kick coming from the accumulation of noise from consecutive single rearrangements. Conditioning $P(x)$ on avalanches $S\approx S_c$ (Fig.~\mbox{\ref{fig5}}b) all but eliminates the intermediate power law ($\tilde{\theta} = 0$), and the pseudogap region immediately gives way to a plateau, whose scaling verifies equation (\ref{eq_for_p}). The onset of the plateau in $P(x \vert S=S_c)$ has the same finite size scaling and therefore corresponds to the plateau in the avalanche kick-distribution $\Delta x_c$. We thus conclude that the first deviation from the pseudogap behavior of $P(x)$ occurs at $x_c\sim \Delta x_c$. }

\bnew{Another scale that could be relevant is the average avalanche size $\langle S \rangle$. This scale plays a central role in the scaling relations for the yielding transition, as it represents the typical stress release occurring during flow, which must (in steady-state flow) be equal to the typical loading. Since the loading between avalanches is controlled by $\xm$, this connects the macroscopic flow to the microscopic description. In the Appendix, we show that this connection manifests in the finite-size scaling of a plateau in the kick-distribution from avalanches of the mean size $S \approx \langle S \rangle$ (Fig.~\ref{fig:si_savg_scaling}(a)).  This plateau coincides with the plateau of $P(x)$ after such avalanches (Fig.~\ref{fig:si_savg_scaling}(b)). However, the rescaling of $P(x)L^p$ vs. $x/\xm$ is not sufficient to obtain a good collapse (i.e.~$x_c \not \sim \xm$) in the pseudogap regime contrary to what has been suggested recently in ref.~\mbox{\cite{Ferrero2020}}.}

\begin{table*}[t]
\centering
\caption{Summary of critical exponents and scaling laws. Error ranges are plausible values given the data. \label{tab:crit_exponents} }
\begin{tabular}{lrrrc}
\hline
Exponent & Defining Eq. & Scaling relation & Measured value & Predicted value\\
\hline
$\theta$ & $P(x>x_c) \sim x^\theta$ & - & $0.52\pm 0.05$ &\\
$c$ & $x_c \sim L^{-c}$ & - & $1.15 \pm 0.05$ &\\
$p$ & $P(x\ll1)\sim L^{-p} $& $p = c\theta$ &  0.61$\pm 0.02$ & 0.57\\
$\tau$ & $P(S) \sim S^{-\tau} G(S/S_c)$ & $2 - \frac{c\theta}{d_f}$ & 1.37$\pm 0.05$ & 1.40\\ 
$d_f$ & $S_c \sim L^{-d_f}$ & $d_f=d-c$ & $0.95\pm 0.1$ & 0.85\\ 
$\alpha$ & $\langle x_{min}\rangle \sim L^{-\alpha}$ & $\alpha = d- p$  & $1.4\pm0.02$ & 1.40\\
$\phi$ & $P(\Sigma;L) \sim L^{-\phi} H(\Sigma / L^{-\phi}) $ &- & 0.88$\pm 0.03$ &\\
\hline
\end{tabular}
\end{table*}

\subsection{Interpretation in a random walk picture} 

The above analysis suggests that the crossover from the pseudogap regime is associated with the stress drop of the largest avalanches that release a plastic stress (or strain)  precisely of the order $\sim L^{-(d-d_f)}$ \cite{SalernoRobbins,Tyukodi2019}. 
 In the following, we develop a physical interpretation based on the picture of an elastoplastic block preforming a random walk in residual stress space in the presence of an absorbing booundary at $x=0$ (local yielding) \cite{LinWyartPRX}. 
 A stable walker located in the pseudogap region is going to feel a stress redistribution originating from the rearrangement of the weakest site destabilized through extremal dynamics. If a plastic rearrangement occurs in a site nearby, the kick felt by the stable walker is going to be of the order of $\delta x_u \sim 1 $ and if the kick is destabilizing, then the walker has a high chance of being absorbed. This situation can occur whatever the size of the avalanche. However, if the rearrangement is happening very far from the walker, the latter will experience only a far field effect, that is very sensitive to the size of the avalanche. In other words, it will only feel a kick proportional to the value of the lower cutoff of the mechanical noise. Obviously, in that case, the size of the avalanche matters. But as the walker is still located at $x>x_c$, it is not absorbed and its exploration can continue. The previously described scenario, i.e absorption triggered mainly by the near field, is valid up to moment where the walker reaches $x \sim \Delta x_c \sim x_c$. For this value of $x$, the largest avalanches induce far field kicks that can trigger the absorption of the stable site. Consequently, for all stable walkers located around and below $x_c$, large avalanches increase the probability of being absorbed. Large avalanches are not predominant, and it is more likely to have smaller avalanches. This makes it possible for the walker to explore the region below $x_c$ but now, the mechanical noise associated with avalanches of smaller size can also trigger absorption. The closer the walker comes to the absorbing boundary, the higher are the chances of absorption as almost all types of avalanches (extended or less extended) can induce a far field kick that can trigger absorption. Actually, the walker can explore small residual stress values up to $x \sim \delta x_c$, the lower cutoff of the stress kicks from the smallest events. In this case, surviving is only possible if the kick is stabilizing. \\
 The enhanced absorption for $x \le x_c$ manifests by an increase of the drift in that region and is associated with an increase of destabilizing kicks related to far field rearrangements. The number of large avalanches increases with the system size, and so does the probability of being absorbed below $x_c$. This explains why $\tilde{\theta}$ is decreasing with $L$ and should completely vanish in the thermodynamic limit. 
 We expect that the departure of the pseudogap regime at $x_c \sim L^{-(d-d_f)}$ is shared by every EPM provided that small enough $x$ values are investigated. Although other recent works showing the departure from the pseudogap regime do not envision this scenario, they report values of crossover exponents $c$ compatible with $c=d-d_f$ both in 2d \cite{Tyukodi2019,Ferrero2020} and in 3d \cite{Ferrero2020}.
 
\subsection{Scaling relations} An important consequence of our findings pertains to the finite size scaling of the weakest sites in the system. Assuming that residual stresses can be seen as independent random variables, extreme value statistics dictates that $P(x)\sim x^{\theta}$, $\langle x_{min}\rangle\sim L^{-\alpha}$ with $\alpha=d/(1+\theta)$. Setting $\langle x_{min}\rangle\sim \langle S \rangle L^d$ leads to an important scaling law linking the pseudogap exponent to the avalanche statistics, $\tau=2 - \theta \alpha / d_f = 2 - \theta d/(d_f(\theta +1))$ \cite{LinWyart2014}.
As shown in the Appendix, this relation continues to hold even in the presence of an intermediate power law or plateau in $P(x)$ but only if the departure vanishes at least as fast as $\xm$ with increasing system size \cite{Ferrero2020}. 
However, the fluctuations of total stress observed in our data indicate that correlations among the random variables play a significant role. We also show in the Appendix that when $P(x)$ gives way to a plateau below $x_c\sim \Delta x_c$,  $\alpha = d-p$ in the thermodynamic limit, and thus with $c=d-d_f$
\begin{align}
    \tau = 2 - \frac{(d-d_f)\theta}{d_f}\, .
\end{align} 
With our measured value of $p=0.61$, $\theta = 0.52$, and $d_f = 0.95$ in the accessible range of system sizes, we find that $\alpha=1.39$ and $\tau=1.43$. These new scaling relations for $\alpha$ and $\tau$ are therefore in better agreement with our values $\alpha=1.4$ and $\tau = 1.37$ than the previously derived scaling laws $\alpha=d/(1+\theta)$ and $\tau = 2-\theta d /(d_f(1+\theta))$ \cite{LinWyart2015} which would predict $\alpha = 1.32$ and $\tau = 1.28$.
Table~\ref{tab:crit_exponents} summarizes all critical exponents and scaling laws proposed in the present work.

\section{Conclusion}
\red{In an amorphous solid sheared under athermal quasistatic conditions, the stress of a local region evolves due to both individual and collective plastic events elsewhere. Despite the presence of large avalanches, the unconditioned distribution of mechanical noise in our EPM implementation is still mostly dominated by small plastic events coming from individual STs bounded from below by the system size. This first characteristic scale manifests as a terminal plateau in the distribution of residual stresses that vanishes as $\sim L^{-d/\mu}$ with increasing system size and becomes irrelevant in the thermodynamic limit. This plateau corresponds to the one discussed in previous works \cite{Ferrero2020}, where it was attributed to the scale set by the typical or mean size of the stress kicks. Our interpretation is compatible with this picture because for $1\le \mu <2$, the mean of the kick distribution is indeed proportional to the lower cutoff which sets the smallest physical scale.} 

Nevertheless, the mechanical noise from extended plastic events has dramatic implications. The stress change due to the largest avalanches sets a second, larger characteristic scale for the magnitude of the mechanical noise that causes a departure from the pure power law form for larger residual stresses. Below the crossover $x_c$, large avalanches become rare as a consequence of the enhanced absorption related to large events.   

\red{For any finite system size $L$, our calculations reveal a previously unnoticed intermediate power-law regime in the distribution $P(x)$ below $x_c$ that is characterized by a system-size dependent exponent $\tilde{\theta}(L)$. The average size $\langle S \rangle$ of the plastic events increases with system size, implying an increase of the drift and an enhanced reduction in the density of individual rearrangements, leading to the plateauing of the intermediate power-law in the thermodynamic limit. Since the crossover is set by large avalanches, we expect $\xm$ to belong to an extended plateau region.  Indeed $\xm$ vanishes more quickly with increasing system size than $x_c$. As explained above, this implies $\xm \sim L^{-\alpha}$ with $\alpha=d-\theta(d-d_f)$  at all system sizes and determines the average value of weakest site which in turn controls the mean avalanche size. In our new scaling relation, the avalanche shape as described by the fractal dimension $d_f$ enters explicitly. This insight is the central result of our work. }

Our results offer possible new routes of interpretation for the yielding transition. In particular, it would be interesting to see how the phenomenology observed here in two dimensions would manifest in three dimensions as the geometry of avalanches encoded in $d_f$ would change. Moreover, one might wonder whether the picture of correlated events inducing the departure of the pseudogap regime is still valid in the transient regime for which recent results from atomistic simulations reported the appearance of a plateau in $P(x)$ after only few percent of deformation  \cite{Ruscher2020} and an increase in the fractal dimension \cite{Oyama2020} with respect to the elastic regime \cite{shang2020elastic}.

\begin{acknowledgements}
We thank Peter Sollich and Jack Parley for a critical reading of our manuscript. This research was undertaken thanks, in part, to funding from the Canada First Research Excellence Fund, Quantum Materials and Future Technologies Program. High performance computing resources were provided by ComputeCanada. C.R acknowledges financial support from the ANR LatexDry project, grant ANR-18-CE06-0001 of the French Agence Nationale de la Recherche.
\end{acknowledgements}


\noindent
\appendix
\section*{Appendix:}


\section{Methods} 
\subsection{Elastoplastic model implementation}
In our elastoplastic model (EPM), we use a finite-element method on a regular triangular mesh to determine the stress propagation between sites and the stress field from applying displacements at the boundaries of the system. Each square site in the system is a plaquette consisting of four edge-sharing triangles. We use first-order Lagrange elements for the displacements $\vc{u}$, which can be understood as setting the displacement at each vertex of the mesh and assuming that the displacement varies linearly within each cell (triangle). As the strain,
\begin{equation}
   \tens{ \gamma  } = \frac{1}{2}\left( (\vc{\nabla} \vc{u}) + (\vc{\nabla} \vc{u})^T \right) \,,
\end{equation}
involves the gradient of Lagrange-1 elements, the natural finite element space for the strain and stress are zeroth order discontinuous Galerkin elements. These elements can be understood as setting a constant tensor within each cell. We divide the total strain, $\tens{\gamma}$ into elastic and plastic (stress-free) contributions, with $\tens{\gamma} = \tens{\gamma_{el}} + \tens{\gamma_{pl}}$. We treat the elastic contribution to strain with linear isotropic homogeneous elasticity, so \begin{equation} 
\tens{\sigma} = 2 \mu  \tens{\gamma_{el}} + \lambda \text{Tr}(\tens{\gamma_{el}})  \mathds{1} \,,
\end{equation}
where $\mu$ and $\lambda$ are the Lam\'e parameters. While irrelevant for critical dynamics, we use $\mu=20$ and $\lambda=10$. 
By combining this with the equilibrium condition $\nabla \cdot \bm{\sigma} = 0$, fixing the displacements of the mesh at the boundaries, and setting the plastic strain on each plaquette, the stress-field is completely determined.



The most expensive part of the simulation is in solving the stress field. We use Python to calculate the stress on each plaquette, and to drive a parallelized FEM solver, FEniCS\cite{logg2010dolfin,alnaes2015fenics}.

\begin{figure}[t]
\centering
    \includegraphics{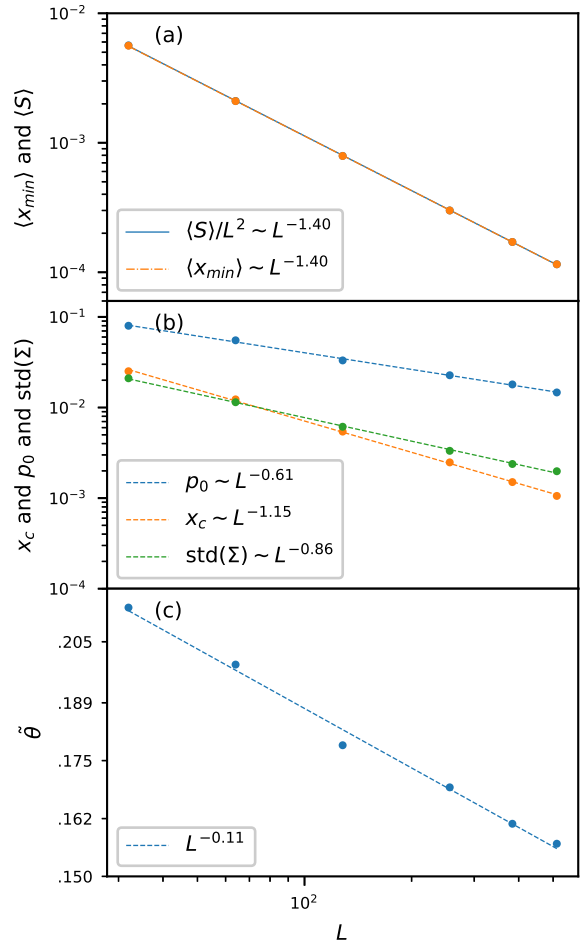}
	\caption{Finite size scaling for (a) the average avalanche size  $\langle S \rangle = N\langle \Delta \Sigma \rangle$ and average proximity to failure $\xm$ after an avalanche. (b) Finite-size scaling for $x_c$ as computed by computing the intersection between the intermediate power-law (estimated by fitting through $\xm$, 	$P(x={\xm})$ and $\delta x_c$, $P(x=\delta x_c)$) and the main power-law $x^\theta$, from Fig.~1 in the main text, along with the terminal plateau value $\lim_{x\rightarrow 0^+} P(x) \rightarrow p_0$, and the standard deviation of the initiation stress $\text{std}(\Sigma)$.  (c) The corresponding $\tilde{\theta}$ finite-size scaling extracted from the procedure.  Dashed lines in (a-d)  are least-square fits to log-log data.}
	\label{fig:fss}
\end{figure} 

\subsection{Calculation of the drift velocity}
The drift velocity is a concept originating in mean-field random-walker models \cite{LinWyartPRX}. To make contact with that literature, for instance, in the prediction for $\tilde{\theta}$ \cite{Doussal2009}, we must define a drift velocity experienced after each elementary kick a site experiences. This is clearly a drift velocity on a per-ST basis. However, our loading protocol introduces a drift ($x_{min}$) to initiate an avalanche, and so the drift velocity is the same for all the STs comprising that avalanche. For an avalanche (labelled $
i$ here) preceded by loading $x_{i,min}$ with $\ell_i$ events, the drift velocity is therefore defined as $N x_{i,min} / \ell_i$. Each of the $\ell_i$ STs (labelled here `j') comprising  avalanche $i$ experiences this drift. Hence, after measuring $M$ avalanches, the average drift velocity on a per-ST basis can be written (summing first over avalanches, and then over their constitutive STs): 
\begin{eqnarray}
\langle v \rangle_{ST} &\equiv& N \langle \frac{x_{min}}{\ell} \rangle_{ST}  = N\frac{1}{\sum_{i=1}^{M} \ell_i}\sum_{i = 1}^M\sum_{j=1}^{\ell_i} x_{i,min} / \ell_i\nonumber \\ 
&=&\frac{N}{\langle \ell \rangle_{av} M}\sum_{i = 1}^M x_{i,min} = \frac{N\langle x_{min} \rangle_{av}}{ \langle \ell \rangle_{av}}
\end{eqnarray}
\bnew{
Thus, we calculate the $\langle v \rangle_{ST}(x)$ in Fig.~\ref{fig3} as  $\langle v \rangle_{ST}(x) = N \frac{\langle x_{min}\rangle (x)}{\langle \ell \rangle(x)}$. The averages $\langle x_{min}\rangle (x)$ and $\langle \ell \rangle(x)$ are determined by, after each avalanche $i$ computing a triplet of values for each site $l$,  $(x_l,x_{min,i},\ell_i)$. Each triplet is binned according to the $x_l$, and then within each bin centred at $x$ the average $\langle x_{min} \rangle$(x) and $\langle \ell \rangle$(x) is computed. }

\section{Finite size scaling of averaged quantities, crossovers, and intermediate plateau exponent}
Figure \ref{fig:fss} reports finite size scaling data for the mean avalanche size 
$\langle S\rangle$, the mean value of the weakest site $\xm$, the crossover from the pseudogap regime $x_c$, the intermediate plateau exponent $\tilde{\theta}$, and the plateau value $P_0$ itself. Also shown is the finite size scaling of std$(\Sigma)$, the standard deviation of the global stress fluctuations about the mean flow stress.

\section{Finite size scaling of kick and residual stress distributions with $\langle S \rangle$ and $\xm$}
\begin{figure}[t]
    \centering
    \includegraphics{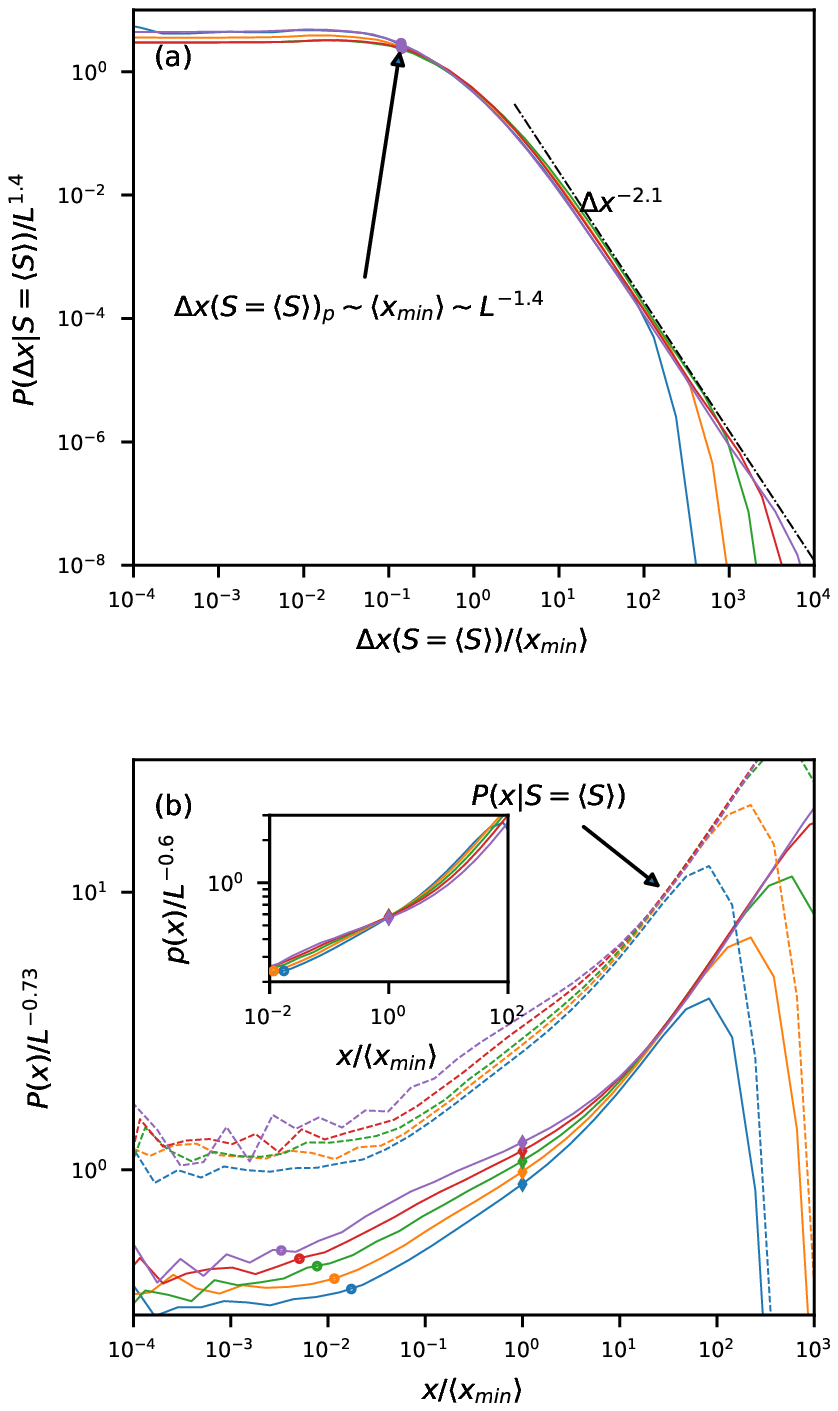}
    \caption{Finite size scaling for (a) the kick distribution for plastic events near to the typical size $S\approx \langle S \rangle$ and (b) for the $P(x)$ and $P(x|S=\langle S \rangle )$ distribution. The $P(x)$ distribution is offset in the y-direction by a factor of $1/3$ for clarity. As in Fig.~\ref{fig1}, diamonds mark $\xm$ and circles mark $\delta x_c$. Inset: FSS collapse for $\xm$, $P(x=\xm)$ fails to collapse the upper-power law, indicating that the crossover $x_c$ from main to intermediate power-law has a different finite-size scaling.}
    \label{fig:si_savg_scaling}
\end{figure}

\bnew{Another scale for avalanche size one can consider is the average avalanche size, $\langle S \rangle$. This average must be equal to the mean stress accumulated during loading $\mu L^2 \langle x_{min}\rangle  = \langle S \rangle$. This equivalence plays a central role in linking the macroscopic flow to the distribution of microscopic weak sites. In Fig.~\mbox{\ref{fig:si_savg_scaling}}a, we see this connection manifest as $L^{-1.4} \sim \xm \sim \Delta x(S=\langle S \rangle)_p$. This plateau in the kick-distribution appears to set the scale for the terminal plateau in $P(x | S\approx \langle  S  \rangle )$ (Fig.~\mbox{\ref{fig:si_savg_scaling}}b). }

\bnew{Therefore, $\xm$ does collapse some features in the $P(x \vert S\approx \langle S \rangle)$
distribution. However, as the inset shows, the points $\xm$, $P(\xm)$ in the unconditioned distribution cannot be simultaneously collapsed with the main $\sim x^\theta$ power-law, indicating that the cross-over from the main power-law $x_c$ does not scale with $\xm$.}

\section{Scaling relations for the yielding transition}

\subsection{Crossover related to spanning plastic events}

Our data suggests that the upper-transition between the $\theta$ and $\tilde{\theta}$ regimes is set by the lower cutoff scale of the kick distribution for the largest avalanches, with $s \approx  s_c \sim L^{0.95}$ implying $\Delta x_c \sim L^{2-d_f} \sim L^{-1.05} > L^{-1.4} \sim \xm$. That is, $\xm$ occurs in the $\tilde{\theta}$ regime. This impacts the finite size scaling. If we assume a residual stress distribution of the form
\begin{equation}
P(x;L)= \left\lbrace
\begin{array}{ll}
	& p_0(L) \hspace{1.0cm} \forall \,  \,  x \le x_p	 \\
    & c_1(L) x^{\tilde{\theta}(L)} \hspace{0.3cm} \forall  \,   \,  x_p	 \le x \le x_c \\
	& c_2 x^{\theta} \hspace{1.2cm} \forall  \,   \,  x \ge x_c
	\label{pdex_eq_redux} 
\end{array}
\right. \,,
\end{equation}
 then the self-consistency equation for $\xm$ is given by:
\begin{equation}
    L^{-d} = \tightint{0}{\xm}P(x;L)\dup x  = p_0 x_p + \frac{c_1}{\tildet + 1 }(\xm^{1+\tildet} - x_p^{1+\tildet} )\,.
\end{equation}
Using continuity of $p(x;L)$ at $x_p$, we obtain $p_0(L) = c_1(L)x_p^{\tildet}$, with which we can eliminate $c_1$ in favour of $p_0$, to obtain after some algebra,
\begin{equation}
    \xm^{1+\tildet} = \left[ \frac{L^{-d}}{p_0} - x_p + \frac{x_p}{1+\tildet} \right] x_p^{1+\tildet} (1+\tildet)  \,.
\end{equation}
Now in the large $L$ limit, the $\tildet$ drop off, and the dominant term is $L^{-d} / p_0 \sim L^{p-d}\sim \xm \sim L^{-\alpha}$, giving the scaling relation $\alpha = d-p$. As outlined in the main text, with $p=c\theta$ and $c = d-d_f$ one obtains $\alpha = d - \theta(d-d_f)$. This scaling relation can also be readily derived by considering the complementary integral, $L^{-d} = 1 - \int^{x_{max}}_{\xm} P(x;L)\dup x$, and applying continuity.

\subsection{Pseudogap or crossover at $\xm$.}
If $x_c$ is smaller than, or scales with, $\xm$, the story is different. Using the ``ideal'' pseudogap form $P(x)\sim x^\theta$ with the self-consistency equation 
\begin{equation}
    L^{-d} = \int_0^{\xm} P(x) dx  \label{eq:scaling_lower_px_dependence}
\end{equation}
yields immediately the scaling  $\xm\sim L^{-\alpha}$ with $\alpha=d/(1+\theta)$ \cite{LinWyartEPL}. Initial inspection would suggest that $\xm$ should depend on the form of  $P(x)$ below $\xm$. As a consequence, one might imagine that the above scaling relation no longer holds. This is, however, not the case. Consider $P(x)$ of the form: 
\begin{equation}
P(x;L) = 
\begin{cases}
0 & x > \xx \\
C(L)x^\theta  & x\ge \xm \text{ and } x<\xx \\
f(x;L) & x< \xm\\
0 & x < 0
\end{cases}
\end{equation}
for a non-decreasing $f(x)$, where $\xm \sim L^{-\alpha}$. Considering the complementary probability, equation~\ref{eq:scaling_lower_px_dependence}, $\xm$ can be written to depend only on the power-law above $\xm$: 
\begin{align*}
L^{-d} &= 1 - \int_{ \xm }^{\xx} Cx^\theta dx  \\ 
    &= 1 - \frac{C}{1+\theta} \left[\xx^{1+\theta} - \xm^{1+\theta}\right]
\tagalign{\label{eq:scaling_upper_px_dependence}} \, .
\end{align*}
The normalization factor, $C$, will necessarily exhibit finite-size scaling.
\begin{equation}
C = \frac{(1+\theta)(1-z^{d})}{\xx^{1+\theta} - \xm^{1+\theta} }
\end{equation} We can expand $C$ as a Taylor series in $ z = \frac{1}{L}$. Then to lowest order,
\begin{equation}
    C = (1+\theta) \left[ \frac{1}{\xx^{1+\theta}} + -\frac{d!}{\xx^{1+\theta}} z^d + \frac{(\alpha(1+\theta))!}{\xx^{2+2\theta}}z^{\alpha(1+\theta)} \right]
\end{equation}
with higher-order terms omitted. Since, \textit{a priori} $\alpha(1+\theta) = t$ could be less than $d$ (but still a  positive integer, so as to ensure that all of terms in the  Taylor series are well defined) we include both $z^d$ and $z^t$ terms. The requirement that $f$ is non-decreasing, along with continuity, ensures that $L^d = \int_0^{\xm} f(x)dx  \le \xm^{1+\theta}$, implies that $t \ge d$. Hence, the lowest order term in the Taylor expansion of $C$ is $\mathcal{O}(z^d)$. Thus, to lowest order, equation~\ref{eq:scaling_lower_px_dependence} gives 
\begin{equation}
    \xm^{1+\theta} \sim L^{-d}
\end{equation}
which gives the scaling relation $\alpha(1+\theta) = d$. \bnew{ In summary, if the cross-over from the main power-law $p(x)\sim x^\theta$ occurs at or below $\xm$, the typical scaling law is restored. In Fig.~\ref{fig:si_savg_scaling}, we test this hypothesis, and find that the curve-collapse for the crossover $x_c$ does not scale in proportion to $\xm$.  }

\subsection{Summary of exponents and scaling laws}
In light of the alteration in the finite-size scaling of $\alpha$, it is reasonable to ask how the scaling relation $\tau = 2-\frac{\theta d}{d_f (1+\theta)}$ is changed. This relation was introduced in \cite{LinWyart2015}, by beginning with 
\begin{equation}
    \langle S \rangle\sim L^{d_f(2-\tau)}
\end{equation}
which comes from assuming $P(S;L) \sim S^{-\tau} G(S/S_c)$ with $S_c \sim L^{d_f}$. Now, $S = - L^{d} \Delta\Sigma $ where $\Delta \Sigma$ is the stress drop over the course of an avalanche. Hence, $\langle \Delta \Sigma\rangle \sim L^{-d + d_f(2-\tau)}$. In the steady state, this is equal to the loading from $\langle x_{min}\rangle \sim L^{-\alpha}$. Hence,
\begin{equation}
    \alpha = d - d_f(2-\tau)
\end{equation}
or 
\begin{equation}
    \tau = 2-(d-\alpha)/d_f. 
\end{equation}
When $x_c > \xm$, we retain the relation $\alpha = d / (1+\theta) $, which implies $\tau =2- d\theta / [d_f(1+\theta)]$. However, in the case that $\alpha = d-p$, we have
\begin{equation}
\tau = 2-p/d_f
\end{equation}
or $\tau = 2 - c\theta / d_f$.

\bibliography{references}

\end{document}